\documentstyle[aps]{revtex}

\def\be {\begin{equation}}
\def\ee {\end{equation}}
\def\ba {\begin{eqnarray}}
\def\ea {\end{eqnarray}}
\def\nn {\nonumber}
\def\a  {\alpha}
\def\b  {\beta}
\def\c  {\gamma}

\def\D  {\Delta}

\def\L  {\Lambda}

\def\O  {\Omega}
\def\p  {\pi}

\def\r  {\rho}
\def\th {\theta}

\def\t  {\tau}
\def\la {\label}
\def\le {\left}
\def\ri {\right}
\def\pa {\partial}
\def\f {\frac}

\def\no {\noindent}
\def\bi {\begin{itemize}}
\def\ei {\end{itemize}}

\def\vs {\vspace}

\def\ul {\underline}

\begin{document}
\draft
\title{General Logarithmic Corrections to Black Hole Entropy}
\author{ Saurya Das }
\address{ Physics Dept., The University of Winnipeg and} 
%\address{}
\address{Winnipeg Institute for Theoretical Physics,}  
\address{515 Portage Avenue, 
%\address{ 
Winnipeg, Manitoba R3B 2E9, CANADA }
\address{ EMail: saurya@theory.uwinnipeg.ca   }
\author{Parthasarathi Majumdar}
\address{ The Institute of Mathematical Sciences,} 
\address{ C.I.T. Campus, Chennai - 600 113, INDIA}
\address{EMail: partha@imsc.ernet.in}
\author{ Rajat K. Bhaduri}
\address{Dept. of Physics and Astronomy,}
\address{Mc Master University,}
\address{ Hamilton, Ontario L8S 4M1, CANADA}
\address{EMail: bhaduri@mcmail.cis.mcmaster.ca } 
%\address{  }
\maketitle

\begin{abstract}

We compute leading order corrections to the entropy of any 
thermodynamic system  
due to small statistical fluctuations around 
equilibrium. When applied to black holes, these corrections are shown 
to be of the form $-k\ln (\mbox{Area})$. For  
BTZ black holes, $k=3/2$, as found earlier. We extend the result
to anti-de Sitter Schwarzschild and 
Reissner-Nordstr\"om black holes in arbitrary dimensions.
Finally we examine the 
role of conformal field theory in black hole entropy and its corrections. 

\end{abstract}

%%%%%%%%%%%%%%%%%%%%%%%%%%%%%%%%%%%%%%%%%%%%%%%%%%%%%%%%%%%%%%%%%%%%%%%%%%%

\section{Introduction}
\la{intro}
Since it is more or less certain that black holes much larger than the
Planck scale have entropy proportional to its horizon area 
\cite{bh,string,ash1,car2,solo}, it is 
important to investigate what the leading order corrections are, as
one reduces the size of the black hole. 
There have been several attempts in this direction. For example, 
in non-perturbative quantum general relativity (NPQGR) (aka Quantum Geometry)
\cite{ash}, 
the density of microstates has been computed for non-rotating asymptotically 
flat (AF) $4d$ black holes \cite{ash1,km}. In this approach, the black hole
horizon is treated as a boundary of spacetime which is quantized within the
NPQGR formalism. The 
dimensionality of Hilbert space of this theory,
which is computed by counting the conformal blocks of a
well-defined $2d$ CFT, namely $2d$ $SU(2)_k$ Wess-Zumino model,
gives the density of
states. For large black holes, the logarithm of this density of 
states is seen to be precisely the Bekenstein-Hawking entropy {\it together 
with corrections :} $-3/2 \ln (\mbox{Area}) $ \cite{qg,qg1}. 
Similar logarithmic corrections were found for quantum Schwarzschild 
black holes in \cite{fursaev} and for BTZ, string theoretic
and all other black holes whose microscopic degrees of freedom
are described by an underlying conformal field theory 
in \cite{carlip} using a corrected version of the asymptotic Cardy formula.
More recently, precisely such corrections have been obtained for
three dimensional BTZ black holes by computing the exact 
partition function \cite{gks}.
Logarithmic corrections to black hole entropy were also obtained
in \cite{other} by considering matter fields in black hole 
backgrounds,  in \cite{solo1} in the context of string-black hole 
correspondence, in \cite{jy} for dilatonic black holes and
in \cite{bss} using Rademacher expansion of the partition function. 
Thus, logarithmic corrections to Bekenstein-Hawking entropy appear to be a 
generic property of black holes. The question is why are they always
logarithmic.  

We try to address this question here. 
In this paper, we show that logarithmic corrections to thermodynamic
entropy arise 
in {\it all} thermodynamic systems when small stable fluctuations
around equilibrium are taken into account. 
The stability condition is equivalent to the specific
heat being positive, so that the corresponding canonical ensemble is stable.
These fluctuations determine
the prefactor in the expression for the density of states of the 
system, whose logarithm 
gives the corrected entropy of the system  (similar to the entropy 
correction due to the exact prefactor in Cardy formula, ref.\cite{carlip}).  
As shown in section (\ref{partition}), 
these corrections are always logarithmic in nature, and when applied to 
black holes in section (\ref{application}), 
they indeed predict the form $-k\ln (\mbox{Area})$. Thus logarithmic 
corrections to Bekenstein-Hawking entropy can be interpreted as 
corrections due to small thermal fluctuations of the black hole around its
equilibrium configuration.  
Note that this prescription applies  to {\it all} black holes
with positive specific heat. Also
note that this analysis simply uses macroscopic black hole properties
such as expressions of entropy and temperature in terms of its mass and 
charge etc, but does not use properties of 
the underlying microscopic theory of gravity (CFT, quantum geometry, 
string theory
etc). In this sense, the corrections are indeed universal, and will
apply to almost all black hole spacetimes, irrespective of whether they
arise in Einstein gravity or some other versions of general relativity, 
whether they are supersymmetric or not.

When applied to BTZ black holes, the prefactor turns out to 
be $k=3/2$, as found in \cite{carlip}. 
Corrections to entropy of AF Schwarzschild black holes 
cannot be found in this way, since the specific heat of the 
latter is negative and its canonical ensemble is unstable \cite{euclidean,hp}. 
However, the problem can be circumvented by considering spherically 
symmetric black holes which are aymptotically anti-de Sitter, and not AF. 
Then, once again, the leading order entropy
corrections are logarithmic in nature. 
Similarly, for Reissner-Nordstr\"om black holes in arbitrary 
spacetime dimensions, there exists a range of mass and electric charge,
for which the specific heat is positive. In this case too, logarithmic
corrections result.  

Finally, in section (\ref{cft}) we examine thermodynamic entropy as  
an exact function of the inverse temperature $\b$ (which is not necessarily 
the equilibrium temperature), namely $S = S(\b)$. We note that the function 
suggested by CFT \cite{carlip}, and suitable generalizations thereof,
give rise to identical logarithmic corrections (including coefficients) 
as found in the earlier section from pure thermodynamic considerations.

It may be noted that in the analysis of this paper, black holes are
still considered large (macroscopic) compared to the Planck scale. 
This ensures that the logarithmic terms are indeed much smaller compared
to the Bekenstein-Hawking term, and can thus be regarded as `corrections'.

\section{ Canonical and Microcanonical Entropy } 
\la{partition}

We consider a canonical ensemble with partition function
\be
Z(\b)
%= \sum_i \r(E_i) e^{-\b E_i}
= \int_0^\infty \r(E) e^{-\b E} dE~~,
\label{part1}
\ee
where $T=1/\b$ is the temperature in units of the Boltzmann constant
$k_B$.    
The density of states can be obtained from (\ref{part1})
by doing an inverse Laplace transform (keeping $E$ fixed) 
\cite{bm,rkb}:
\be
\r(E) = \f{1}{2\p i} \int _{c-i\infty}^{c+i\infty} Z(\b) e^{\b E} d\b
= \f{1}{2\p i} \int_{c-i\infty}^{c+i\infty} e^{S(\b)}d\b~~,
\la{density1}
\ee
where
\be
S(\b) = \ln Z(\b) + \b E
\label{basic}
\ee
is the {\it exact} entropy
as a function of temperature, not just its value at equilibrium.
It is formally defined as the sum of entropies of subsystems of
the thermodynamical system, which are small enough to be themselves in
equilibrium. The complex integral can be performed by the method of steepest
descent around the saddle point $\b_0 (= 1/T_0)$, such that
$S'_0 := (\pa S(\b)/\pa \b)_{\b=\b_0}=0$. $T_0$ is the equilibrium
temperature, such that the usual equilibrium relation
$E=-(\pa \ln Z(\b)/\pa \b)_{\b=\b_0}$ is obeyed. Expanding $S(\b)$ about
$\b=\b_0$, we get
\be
S = S_0 + \f{1}{2} (\b - \b_0) ^2 S_0'' + \cdots ~~,
\label{ent1}
\ee
where $S_0 := S(\beta_0)$ and $S''_0 := (\pa^2 S(\b)/\pa \b^2)_{\b=\b_0}$.
Substituting (\ref{ent1}) in (\ref{density1}) :
\be
\r(E) = \f{e^{S_0}}{2\p i} \int_{c-i\infty}^{c+i\infty}
e^{1/2 (\b - \b_0)^2 S_0''} d\b ~.
\ee
By choosing $c=\beta_0$, and putting $(\beta-\beta_0)= i x$, where $x$ is
a real variable, we obtain, for $S''_0>0$,
\be
\r(E) = \f{e^{S_0}}{\sqrt{2\p S''_0}} ~~.
\la{corr0}
\ee
Note that the density of states $\r(E)$ and $S_0''$ have dimensions of inverse
energy and energy squared respectively. Henceforth, 
we set the Boltzmann constant $k_B$ to unity.
The logarithm of the density of states $\r(E)$ is then the microcanonical
entropy 
\be
{\cal S} := \ln \r(E) = S_0 - \f{1}{2} \ln S_0'' + ~\mbox{(higher order terms)}.
\la{corr1}
\ee
Here ${\cal S}$ is the corrected microcanonical entropy {\it at equilibrium},
obtained by incorporating small fluctuations around thermal equilibrium.
This is to be distinguished from the function $S(\b)$, which is the
entropy at any temperature (and not just at equilibrium).
The above formula is applicable to all thermodynamic systems considered
as a canonical ensemble, including black holes. For the latter,
$S_0$ is the Bekenstein-Hawking area law. It is seen that the
correction solely depends on the quantity $S_0''$. Moreover, the second
term in Eq.(\ref{corr1}) appears
with a negative sign, as conjectured in general in \cite{qg1} from
the holographic point of view. Now, we will
estimate $S_0''$, without assuming any specific form of $S(\b)$.
From Eq.(\ref{basic}), it follows that
\be
S''(\b)={1\over Z}({\pa^2 Z(\b)\over {\pa \b^2}})-
{1\over Z^2}({\pa Z\over \pa \b})^2~.
\ee
This means that $S_0''$ is nothing but the fluctuation squared of energy
from the equilibrium, i.e., 
\be
S_0''= <E^2>-<E>^2~, 
\ee
where, by the definition
of $\b_0$, $E=<E>= - (\pa \ln Z/\pa \b)_{\b=\b_0}$. It immediately follows
that
\be 
S_0''= T^2 C
\label{esci}
\ee
where $C \equiv (\partial E/ \partial T)_{T_0}$ is the dimensionless
specific heat. The above result also follows from a detailed 
fluctuation analysis of a stable thermodynamic system \cite{ll}.
 
Substituting for $S_0''$ from (\ref{esci}) in (\ref{corr1}), we get:
\be
{\cal S} = \ln \r = S_0 - \f{1}{2} \ln \le( C~T^2 \ri) + \cdots
\la{corr3}
\ee
 
This is our master formula, and for dimensional correctness, it is understood
that the quantity within the logarithm is divided by $k_B^2$.
As mentioned earlier, the above correction arises for any thermodynamic
system satisfying the first law. We would like to
apply this to the particular case of black holes \cite{brown}. 
To do this, we would replace $T \rightarrow  T_H$, the Hawking temperature. 
Then we calculate $C$ for the specific black holes.
 
Before that, let us determine the limit of validity of the formula
(\ref{corr3}). We assume the quantum fluctuations of the thermodynamic
quatities under consideration, e.g. $S, \b, \cdots$ are small. Denote
such a generic variable by $y$. Then the above restriction translates to
$\D y \ll y$. Then from the
uncertainty relation for quasi-classical systems \cite{ll2}
$$ \D E \D y \geq  {\dot y} $$
and approximating ${\dot y} = y/\t$ (where
$\t$ is relaxation time scale), it follows that:
$$ \D E >> 1/\tau ~~.$$
Then, using $\D E = T\D S$, we get:
$$ \D S >> \f{1}{\t T}~~.$$
Finally, assuming $\D S \ll 1$, we get from above \cite{ll}:
\be
 T >> \f{1}{\t}
\la{limit}
\ee
which sets the limit that (\ref{corr3}) ceases to be valid for very low
temperatures, lower than a characteristic value $1/\t$. For
black holes, $\t$ is presumably of order Planck length, the only
natural length (or time) scale relevant to the system. Note that
the above inequality shows that for black holes very close to
extremality ($T \rightarrow 0$), the fluctuation analysis ceases to
be valid due to large quantum fluctuations.
This is consistent with earlier results which showed that
near extremal black holes are highly quantum objects,
irrespective of their size \cite{bdk}.

\section{Application to Black Holes Entropy} 
\la{application}

\subsection{BTZ Black Hole:}

\vs{.3cm}
\no
First let us consider a non-rotating  
BTZ black hole in $3$-dimensions with metric \cite{btz}:
\be
ds^2 = - \le( \f{r^2}{\ell^2} - 8G_3 M \ri) dt^2
+ \le( \f{r^2}{\ell^2} - 8G_3 M \ri)^{-1}  dr^2
+ r^2 d\th^2 ~~.
\ee
Its Bekenstein-Hawking
entropy and Hawking temperature are given by 
\ba
S_0 &=& \f{2\p r_+ }{4 G_3}  \la{btz3} \\
T_H &=& \f{r_+}{2\p \ell^2} = \le[ \f{G_3}{\p^2\ell^2}\ri] ~S_0
\la{temp3}
\ea
where $r_+ = \sqrt{8G_3 M} \ell$ is horizon radius 
($G_3=3$-dimensional Newton's constant), $M$ being mass of the 
black hole and $\ell$ is related to the cosmological constant by: 
$\L =-1/\ell^2$. 
The first law in this case is:
\be
dM = T_H dS_0 ~ 
\label{firstlaw} 
\ee
Since there are no work terms, the specific heat in this case is simply
a total derivative :
%{}From the above relations, it follows that 
%
\be
C = \f{dM}{dT_H} = S_0 = \le[ \f{\p^2 \ell^2 }{G_3} \ri]~T_H 
%\Rightarrow T_H &=& \le[ \f{G_3}{\p^2 \ell^2 }\ri]~S_0
\la{temp4}
\ee
Plugging in (\ref{temp3}) and (\ref{temp4}) in (\ref{corr3}), we get. 
\ba
{\cal S} = \ln \r &=& S_0 - \f12 \ln \le(S_0  S_0^2\ri) + \cdots \nn \\
&=& S_0 - \f{3}{2} \ln S_0 + \cdots~ \la{btz4}.
\ea
The quantity in square brackets in (\ref{temp3})
is a fixed constant independent of black hole parameters,
and hence can be ignored for $S_0 >> 1$ (macroscopic black holes). 
First, note that the correction term is proportional to the 
logarithm of horizon area. Secondly, the proportionality constant 
$-3/2$ is exactly that found in \cite{carlip}. As claimed earlier, 
this shows that logarithmic corrections of this type can be interpreted
as those due to thermal fluctuations of the black hole around its 
equilibrium. 

Introduction of angular momentum is straightforward. It can be shown that
for rotating BTZ black holes far away from extremality, the formulae
(\ref{temp3}) - (\ref{temp4}) are modified as :
\ba
T_H &=& \le[  \f{G_3}{\p^2 \ell^2} \ri] S_0 
+ {\cal O} \le( \f{J^2}{M^{3/2}}  \ri) 
\nn \\
C &:=& \le( \f{\pa T_H}{\pa M} \ri)_{J}^{-1}  
= S_0 + {\cal O} \le(  \f{J^2}{M^{3/2}} \ri) 
\nn
\ea
where $J$ is the angular momentum, and the 
specific heat $C$ is now calculated keeping $J$ fixed. Thus the entropy is, 
\be
{\cal S} = \ln\r = S_0 - \f{3}{2}\ln S_0 
- \f{(J^2/M^{3/2})}{2S_0} + \cdots 
\ee
Thus to leading order the correction
(\ref{btz4}) is reproduced, and sub-leading angular momentum 
dependent terms contribute further polynomial corrections of order
$1/S_0$, the inverse of Bekenstein-Hawking entropy. Such smaller polynomial
corrections were also predicted in \cite{qg,qg1} in the context of 
quantum geometry. For near-extremal 
(very low temperature) situations, the corrections of above type are 
no longer valid for reasons stated at the end of the last section, 
as the necessary condition (\ref{limit}) is violated. 
 
\subsection{Anti-de Sitter Schwarzschild:}
 
\vs{.3cm}
\no
First let us consider a $d$-dimensional Schwarzschild black hole with metric:
\be
ds^2 = - 
\le( 1 - \f{16\p G_d M}{(d-2)\O_{d-2}r^{d-3}}  \ri) dt^2
+ \le( 1 - \f{16\p G_d M}{(d-2)\O_{d-2}r^{d-3}}  \ri)^{-1} 
dr^2 + r^2 d\O_{d-2}^2 ~~,
\ee
(where $G_d=d-$ dimensional Newton's constant, $d\O_{d-2}^2$ is the
metric on unit $S^{d-2}$ and $\O_{d-2}$ is the area of that unit sphere).
Its entropy and temperature are : 
\ba
S_0 &=& \f{\O_{d-2}~r_+^{d-2} }{4 G_d} \la{scent2}  \\
T_H &=&  \f{(d-3)}{4\p r_+}  \la{sc1}
\ea
The mass is related to the horizon radius as :
\be
M = \f{(d-2)(d-3)\O_{d-2}}{16\p G_d}~r_+^{d-3}
\la{scmass}
\ee
with which one can write
\be
T_H = \le( \f{d-2}{4\p} \ri)^{(d-2)/(d-3)}
\le(  \f{\O_{d-2}}{4G_d M} \ri)^{1/(d-3)}~~. 
\la{sctemp4}
\ee
Again, the first law (\ref{firstlaw}) is obeyed, and
\ba
C &=& \f{dM}{dT_H} = - (d-2) S_0 \label{sccv}
\ea
Since the specific heat is negative, one cannot directly apply
the fluctuation correction formula (\ref{corr3}) to the Schwarzschild
black hole. 
One way to circumvent this problem is to consider anti-de Sitter 
Schwarzschild black holes, whose asymptotic structure is that of 
anti-de Sitter space, instead of asymptotically flat Schwarzschild 
black holes. We will see that for black holes with large
masses, the presence of 
a small negative cosmological constant $\L$ is sufficient to 
make the specific heat positive and render (\ref{corr3}) applicable. 
The microcanonical ensemble for the Schwarzschild 
black hole is of course well defined, and we will return to this issue later.

We parametrize the cosmological constant in $d$-dimensions in terms of
the length scale $\ell$ as: 
$\L = -(d-1)(d-2)/2\ell^2$. The corresponding Einstein's equations admit
of black holes solutions 
%of mass $M$ 
which are asymptotically anti-de Sitter and whose metric is given by:
\be
ds^2 = - 
\le( 1 - \f{16\p G_d M}{(d-2)\O_{d-2}r^{d-3}} + \f{r^2}{\ell^2} \ri) dt^2
+ \le( 1 - \f{16\p G_d M}{(d-2)\O_{d-2}r^{d-3}} + \f{r^2}{\ell^2} \ri)^{-1} 
dr^2 + r^2 d\O_{d-2}^2 ~~.
\ee
Here the parameter $M$ is the conserved charge associated with the 
timelike Killing vector (`mass') \cite{ash2}.
In terms of the horizon radius $r_+$, 
the entropy formula is same as (\ref{scent2}), whereas
the Hawking temperature is :
\be
T_H = \f{(d-1) r_+^2 + (d-3)\ell^2}{4\p \ell^2 r_+}
= \le[ \f{d-3}{4\p} \le( \f{\O_{d-2}}{4G_d} \ri)^{1/(d-2)} \ri]~
\le[ 1 + \f{(d-1) r_+^2}{(d-3) \ell^2}     \ri]~
S_0^{-1/(d-2)} ~~.
\la{adstemp1}
\ee
Although the function $r_+=r_+(M)$ is not explicit,
$C$ can still be calculated using : 
$$ C = \f{(d-2) \O_{d-2} r_+^{d-3}}{4 G_d} 
\le[ \f{\pa \ln T_H }{ \pa r_+}  \ri] ~~~,$$
which yields \cite{mann} :
\be
C =  \le[  (d-2) \f{(d-1)r_+^2/\ell^2 + (d-3) }{ (d-1)r_+^2/\ell^2 
- (d-3) } \ri]~S_0 ~~.
\la{adscv1}
\ee
First, note that the limit $\ell \rightarrow \infty$ reproduces the 
(negative) Schwarzschild value (\ref{sccv}) of $C$. However, for
\be
 \ell < {r_+}{\sqrt{\f{d-1}{d-3}}}~~~, 
\la{ineq1}
\ee
$C$ changes signature and becomes
positive. In the limit of large $\L$ ($\ell \rightarrow 0$), 
$C$ approaches a positive finite value:
\be
C  = (d-2) S_0~~. 
\la{adscv2}
\ee
Consequently, the canonical ensemble becomes stable against 
small fluctuations around equilibrium \cite{hp}. In other
words, introduction of a negative cosmological constant remedies the otherwise 
unstable Schwarzschild canonical ensemble. 

Let us calculate the leading order corrections in this limit, although
one can do so for any $\ell$ satisfying (\ref{ineq1}). From 
(\ref{corr3}), (\ref{adstemp1}) and (\ref{adscv2}), we get:
\be
{\cal S} = S_0 - \f{d}{2(d-2)} \ln S_0  + \cdots  
\la{scadscorr2}
\ee
Thus, once again the leading corrections are logarithmic. 
It is clear that logarithmic corrections result for all values
of $\ell$ lying in the range (\ref{ineq1}). 
Also note from (\ref{ineq1}) that for black hole masses much greater
than the Planck mass, $r_+$ is huge and the range of $|\L|$ 
corresponding to $C>0$ is given by the expression:
\be
|\L| > \f{(d-2)(d-3)}{2 r_+^2}~\approx 0~~. 
\la{range1}
\ee
This means that the introduction of a very small negative 
cosmological constant is sufficient to make the present formalism applicable.

\subsection{ Reissner-Nordstr\"om :}

\vs{.3cm}
\no
For a $d$-dimensional Reissner-Nordstr\"om black hole with metric
\cite{mp}
\ba
ds^2 &=& - \le(  1 - \f{16\p M}{(d-2) \O_{d-2} r^{d-3}}  
+ \f{16\p Q^2}{(d-2)(d-3) r^{2(d-3)}} \ri) dt^2 \nn \\
&+& \le(  1 - \f{16\p M}{(d-2) \O_{d-2} r^{d-3}}  
+ \f{16\p Q^2}{(d-2)(d-3) r^{2(d-3)}} \ri)^{-1} dr^2
+ r^2 d\O_{d-2}^2~~,
\ea
the entropy and temperature are given by:
\ba
S_0 &=& \f{\O_{d-2}~r_+^{d-2} }{4 G_d}  \\
T_H &=&  \f{(d-3)}{2\p~r_+^{d-2} }~
\sqrt{ \le( \f{8\p G_d M}{(d-2)\O_{d-2}} \ri)^2 - 
\f{2G_dQ^2}{(d-2)(d-3)}  }  \la{sc5}
\ea
where $Q$ is the electric charge. 
Using the formula for horizon radius:
\be
r_+^{d-3} = \f{8\p G_d M}{(d-2) \O_{d-2}} +
\sqrt{ \le( \f{8\p G_d M}{(d-2) \O_{d-2}} \ri)^2  - \f{2G_d Q^2}{(d-2)(d-3)} }
~~, 
\ee
the specific heat is :
\be
C =  \le( \f{d-2}{d-3} \ri) S_0~F_2(M,Q) \la{rncq} 
\ee
and the temperature far away from extremality ($r_+ >> r_-$): 
\be
T_H = \le[ \f{d-3}{4\p} \le( \f{\O_{d-2}}{4G_d} \ri)^{1/(d-2)} \ri]~
S_0^{-1/(d-2)} ~F_1(M,Q) \la{rntemp} 
\ee
where
\ba
F_1 (M,Q) &=& 1 - 2G_d \le(  \f{d-2}{d-3} \f{\O_{d-2}Q}{16\p G_d M} \ri)^2 
\nn \\
F_2 (M,Q) &=& \f{\sqrt{  \le(\f{8\p G_d M}{(d-2) \O_{d-2}}\ri)^2 
- \f{ 2G_d Q^2}{ (d-2)(d-3)}} }
{ \f{8\p G_dM}{(d-2)\O_{d-2}} - \f{(d-2)}{(d-3)}
\sqrt{  
\le( \f{8\p G_d M}{(d-2) \O_{d-2} }\ri)^2 - \f{2G_d Q^2}{(d-2)(d-3)} }}~~.  
\nn
\ea
When the electric charge is very small, constant $F_1$  
is of order unity, $F_2$ is negative
and $C$ is close to the (negative) Schwarzschild 
result (\ref{sccv}), i.e. $C \approx - (d-2) S_0$. However, in the range: 
\be
\le[ \f{2(d-2) \O_d^2}{(8\p)^2(d-3) G_d}  \ri] Q^2 <   M^2 
< \le[ \f{2(d-2)^3 \O_d^2}{(8\p)^2(2d-5) G_d}  \ri] Q^2~~, 
\ee
$C >0$ and the leading order entropy correction from (\ref{corr3}) is: 
\be
{\cal S} = S_0 - \f{(d-4)}{2(d-2)} \ln S_0 + \ln \le( F_1^2 F_2 \ri)
+ \cdots  ~~. 
\la{rnent}
\ee
It is interesting to note that the fluctuation corrections in this case
vanish for $d=4$.

As for BTZ, close to extremality, $T_H \approx 0$, 
and the assumptions made in the thermodynamic analysis breaks down, 
as can be seen from (\ref{limit}). Thus (\ref{corr3}) ceases to be valid too. 

Since the string theoretic black holes in $5$ and in $4$ dimensions, whose 
entropy was reproduced by $D$-brane state counting, fall into the 
class of Reissner-Nordstr\"om black holes, we do not consider them
separately.

\section{CFT and Exact Entropy Function $S(\b)$ :}
\la{cft}

\vs{.3cm}
\no
Thus far, we have seen that logarithmic leading order corrections to 
entropy are generic to all statistical mechanical systems near its
equilibrium due to small thermal fluctuations. The quantity $S_0''$ was
sufficient to calculate these corrections, and the former can be
expressed in terms of known thermodynamic quantities $C$ and $T$. 

In other words, an exact entropy function $S(\b)$, although would
be sufficient to get above corrections, was not at all necessary. 
In this section, we will attempt to reproduce these corrections
by postulating an {\it exact} entropy function $S(\b)$, which follows from
CFT and may also perhaps follow from other quantum theories of gravity. 
Thus, on the one hand, it will provide a consistency check between the 
derivations of the previous sections and corrections derived from certain 
microscopic quantum gravitational principles, and on the other hand, it 
will provide an indication as to what kind of entropy functions 
can be expected from underlying theories of quantum gravity.

Let us return to  Eq. (\ref{corr1}), which as mentioned
before, predicts leading correction to the equilibrium entropy in terms
of $S_0''$, the second derivative of the {\it exact} entropy function
$S(\b)$, evaluated at the equilibrium temperature $\b_0$. As stated
before, if we wish to
specialize to black holes, we must first make the following identifications:
\ba
\b_0 &=& \f{1}{T_H} ~~, \\
S_0 = S(\b_0)  &=& \f{A_H}{4G_d}  
\ea
where the first relation follows from the assumption that the black hole is
at equilibrium at Hawking temperature. 
It is clear that this set of information is inadequate to determine
the functional form of $S_0''$, which is independent of both
$S_0$ and $\b_0$.
Geometrically, the quantities $\b_0$ and $S_0$ represent the abscissa
and ordinate of the extremum of the parabola which approximates
the actual curve near extremum. The slope $S_0''$ on the other hand is
still an independent quantity.

To determine $S_0''$, we may assume any specific form
of the function $S(\b)$, which can be well approximated by the
parabolic form (\ref{ent1}). 
The only stipulation is that such a form must admit of an extremum 
at some value $\b_0$ of $\b$. 
If we assume an underlying CFT, then from modular invariance of the partition 
functions of $(1+1)$-dimensional
conformal field theories it follows that $S(\b)$ 
is of the following form (see argument of
exponential function in Eq(2.7) in \cite{carlip} ):
\be
S(\b) = a\b + \f{b}{\b}
\la{cft1}
\ee
where $a,b$ are constants. 
However, to test whether this form is absolutely essential, and whether 
more general functional forms affect results significantly, 
we will assume a more general form which also admits of a saddle
point, and hence the thermodynamic limit. We choose such a function:
\be
S(\b) = a\b^m + \f{b}{\b^n}
\la{ent2}
\ee
wth $m,n,a/b > 0$. 
The form dictated by CFT is a special case ($m=1=n$) of the above. 
Later we will show that (\ref{ent2}) is also consistent with 
assumptions made in the context of Euclidean black hole thermodynamics
\cite{euclidean}. 
A constant additive term that could have been included may be absorbed
in the definition of $S_0$. The above has an extremum at 
\be
\b_0 =  \le( \f{nb}{ma} \ri)^{{1}/(m+n)}  \equiv  \f{1}{T_H}~~~.
\la{temp1}
\ee
Expanding around $\b_0$, by computing the second derivative, we get:
\be
S(\b) = \a~\le(a^n b^m \ri)^{1/(m+n)} + \f{1}{2}~\c
\le( a^{n+2}b^{m-2} \ri)^{{1}/(m+n)}~(\b - \b_0)^2 + \cdots ~,
\ee
where    
\ba
 \a &=& (n/m)^{m/(m+n)} + (m/n)^{n/(m+n)}  \nn \\
\c &=& (m+n) m^{(n+2)/(m+n)} n^{(m-2)/(m+n)} \nn
\ea
are constants.
Comparing with (\ref{ent1}), we find: 
\ba
S_0 &=& \a~\le(a^n b^m \ri)^{{1}/(m+n)} ~~,\\
S_0'' &=& \c~\le( a^{n+2}b^{m-2} \ri)^{1/(m+n)}~~.
\ea
Next, we invert these equations to get $a,b$ in terms of $S_0$ and
$S_0''$ :
\ba
a &=& \f{\a^{(m-2)/2}}{\c^{m/2}}~\le( S_0''\ri)^{m/2}
S_0^{-(m-2)/2} \\
b &=& \le[ \f{\a^{(n+2)/2}}{\c}  \ri]^{-n/2}
S_0^{(n+2)/2} \le( S_0'' \ri)^{-n/2}~~~.
\ea
These values of $a$ and $b$ can now be substituted in
(\ref{temp1}) to obtain:
\be
\b_0 = \f{1}{T_H} = \le( \f{n}{m} \ri)^{1/(m+n)} \sqrt{\f{S_0}{S_0''}~
\f{\c}{\a}}~~~.
\ee
The last relation can be inverted to 
express $S_0''$ in terms of $S_0$ and $T_H$ :
\be
S_0'' = \le[ \le(\f{\c}{\a}\ri)~
\le(\f{n}{m}\ri)^{2/(m+n)} \ri] S_0 T_H^2~.
\ee
This is the relation we sought, since it completely specifies
$S_0''$ (and hence entropy corrections) in terms of Bekenstein-Hawking entropy
and Hawking temperature, and is a direct consequence of our
assumed form of exact entropy function (\ref{ent2}).
Again, the factors in square brackets are independent of black hole
parameters and can be dropped.  
Substituting in (\ref{corr1}), we get:
\be
{\cal S} = \ln \r = S_0 - \f{1}{2} \ln \le( S_0 T_H^2 \ri) + \mbox{smaller terms}
\la{corr2}
\ee
This is the generic formula for leading order corrections to
Bekenstein-Hawking formula. We note that the leading order 
correction is independent of $m,n$ and the latter has an effect only 
in the sub-leading terms. We will return to this issue later.
Now let us apply (\ref{corr2}) to specific black holes. 

\vs{.5cm}
\no
\ul{BTZ :}

{}From (\ref{temp4}), we see that $S_0 = C$ for BTZ. Thus, the 
correction (\ref{corr2}) becomes identical to (\ref{corr3}) ! 
Or, by explicit calculation using  
the relation (\ref{temp3}) between temperature and entropy
of BTZ black hole, we can write (\ref{corr2}) as:
\ba
{\cal S} = \ln \r &=& S_0 -  \f{1}{2}  \ln \le( S_0 S_0^2 \ri) + \cdots  \nn \\
&=& S_0 - \f{3}{2} \ln S_0 + \cdots
\la{btz1}
\ea
and get back the logarithmic correction for BTZ black hole
entropy found in \cite{carlip} and by the general thermodynamic analysis 
(\ref{btz4}), {\it including} the right coefficient
$-3/2$ in front. Thus, the general 
form of the function $S(\b)$ in (\ref{ent2}) is
sufficient to arrive at this correction, and not just the specific
$m=1=n$ case as dictated from CFT. This makes the correction quite robust.
 
As before, angular momentum can be introduced quite easily, and the 
results continue to be valid so long as the black hole is away from 
extremality.

\vs{.5cm}
\no
\ul{ AdS-Schwarzschild :}

Again for Schwarzschild-anti-de Sitter black holes, in the positive 
specific heat domain (\ref{ineq1}), it follows from 
(\ref{adscv1}) that $C = (d-2) S_0$ to leading
order. Then, from (\ref{corr2}), we get:
\be
{\cal S} = \ln \r = S_0 - \f{1}{2} \ln \le(C T_H^2 \ri) +  \cdots
\ee
which agrees with the fluctuation correction formula (\ref{corr3}).
Hence, the logarithmic correction term will have identical coefficient as
in (\ref{scadscorr2}).

\vs{.5cm}
\no
\ul{Reissner-Nordstr\"om :} 

The generalization to charged black holes is also straightforward. 
{}From (\ref{rncq}), we see that $C \approx S_0 F_2(M,Q)$. Thus, from
(\ref{corr2}) we get:
\be
{\cal S} = \ln \r = S_0 - \f{1}{2} \ln \le(C T_H^2 \ri) + \ln F_2 + \cdots
\ee
which is identical to (\ref{rnent}) to leading order.

\section{Discussions}

In this paper, we assumed that a black hole is a 
thermodynamic system in equilibrium at Hawking temperature, and systematically 
analyzed its small statistical fluctuations around equilibrium. These 
fluctuations give rise to a non-trivial multiplicative factor in front
of the expression for the density of states, whose logarithm gives the 
corrected entropy formula. The latter consists of the zeroth order term
which is the Bekenstein-Hawking entropy $A_H/4G_d$ and an additional 
correction term which is logarithmic for all black holes. This shows that
logarithmic corrections to Bekenstein-Hawking entropy arise due to
statistical fluctuations around thermal equilibrium of the black hole.  
This also explains as to why leading correction terms have 
always been found to be logarithmic. Next, we explicitly 
evaluated this logarithmic correction term for various black holes.

Finally, we showed that identical correction terms result if one
{\it assumes} a certain form of the 
entropy function $S(\b)$ which is valid away from 
equilibrium, and which is consistent with conformal field
theory. This consistency on the one 
hand lends additional support to certain underlying theories 
of black hole dynamics that have been used previously and on
the other hand gives an 
indication of the nature of entropy function expected from other 
microscopic theories. 

Now, it may be mentioned that to demonstrate the 
instability of Schwarzschild black holes, it was assumed in 
\cite{euclidean} that the exact entropy function $S(\b)$ is 
obtained from the well known relation at equilibrium (i.e. the 
area law) $S(\b_0)$, by simply substituting $\b_0 \rightarrow \b$.
While this is sufficient for the area law to be valid at Hawking 
temperature, it is certainly not necessary. However, here we will show
that the above assumption is perfectly consistent with the entropy 
function (\ref{ent2}) that we chose in the last section. 

Again consider BTZ black holes. Using (\ref{temp3}) and $\b_0 = 1/T_H$,
we get at equilibrium :
\be
S (\b_0 ) = \f{\p^2\ell^2}{G_3} \f{1}{\b_0}~~.
\ee
Following \cite{euclidean} if we assume that the above continues to hold
for any $\b$, then we get :
\be
S (\b ) = \f{\p^2\ell^2}{G_3} \f{1}{\b}~~.
\la{btzapprox5}
\ee
Also from (\ref{temp3}) and the expression 
$r_+ = \sqrt{8G_3M}\ell$, we can write: 
\be
\b_0 = \f{ \p\ell}{ \sqrt{2G_3 M}}~.
\ee
This implies that for BTZ black hole mass much greater than the Planck mass,
$1/\b_0 \gg \b_0$. Also for $M/M_{Pl} \rightarrow \infty$, let us assume that
the above inequality holds for temperatures slightly away from equilibrium 
(since for truly macroscopic black holes, fluctuations are expected to be
small). Then, from Eq.(\ref{ent2}) it follows that :
\be
\lim_{M/M_{Pl} \rightarrow \infty} S(\b) = \f{b}{\b^n} ~~.
\la{btzapprox6}
\ee
Comparing (\ref{btzapprox5}) and (\ref{btzapprox6}) we find:
\ba
n&=&1 \nn \\
\& ~~~~ b &=& \f{(\p \ell)^2}{G_3} ~~.\nn
\ea
In other words, the assumed relation (\ref{btzapprox6}) is perfectly consistent
with the our general chosen form (\ref{ent2}) in the limit of macroscopic
black holes. This can also be seen from the fact that if (\ref{btzapprox6})
holds with $n=1$, 
then it implies $S_0'' \sim \b^{-3} \sim S_0^3 $, which using (\ref{corr1}) 
in turn implies the required corrected entropy 
$$ {\cal S} = S_0 - \f{3}{2} \ln S_0 + \cdots $$

Similar conclusions follow for AdS-Schwarzschild black hole in $d$-dimensions.
In this case, for large black holes, it again follows from (\ref{ent2}) that 
$S(\b) \approx b/\b^n$. On the other hand, using (\ref{adstemp1}) and assuming
as before that the formula continues to be valid away from equilibrium,
it follows that:
\be
S (\b) = 
\le( \f{4\p \ell^2}{d-1}  \ri)^{d-2} \f{\O_{d-2}}{4G_d} \b^{-(d-2)}~~.
\ee
Comparing, we find: 
\ba
n  &=&  d-2 \nn \\
\&~~~~b &=& \le( \f{4\p \ell^2}{d-1}  \ri)^{d-2} \f{\O_{d-2}}{4G_d} \b^{-(d-2)}
~~.
\nn 
\ea
This shows that the more general form (\ref{ent2}) with $m,n \neq 1$ 
is relevant for black holes other than BTZ.  
The above also implies $S_0'' \sim \b_0^{-d} \sim S_0^{d/(d-2)} $ and 
$$ {\cal S} = S_0 - \f{d}{2(d-2)} \ln S_0 + \cdots $$ 
as found earlier in (\ref{scadscorr2}).

The above examples lend support to the entropy function (\ref{ent2}).
However, since the exponents $m,n$ have no effect on leading 
order corrections (implying the CFT case $m=n=1$ is sufficient), 
it is tempting 
to speculate that conformal field theory underlies all black hole
descriptions in nature as was proposed in \cite{car2}. 

Finally, let us comment on the role of the current analysis to 
entropy of $d=4$ Schwarzschild black holes. The log corrections found in
\cite{qg,qg1} clearly does not fall into this category, since as
remarked earlier, the instability
of canonical ensemble for Schwarzschild black holes 
implies that 
its corrections should be calculated in the microcanonical picture, whereby
the energy (and hence area) of the black hole is held fixed. 
This is precisely what was done in \cite{qg,qg1}, where
the area was held fixed during counting of the dimension of Hilbert space of
states of spin networks. Presumably, similar analysis should be done for
all black holes which have negative specific heat. Thus, the 
computation done here 
and that in \cite{qg,qg1} actually complement 
each other and show that logarithmic corrections are indeed universal.
Since stability of a Schwarzschild black hole can also be
ensured by confining it in a box \cite{by}, it would be interesting to see
what the entropy corrections are in that case.  
Also, one can try to apply our corrections after taking into account
high temperature degrees of freedom, in which case the specific 
has been claimed to be positive \cite{gm}.
It may be noted that the entropy correction found
here has the interesting property
$$ {\cal S} ( A_1 + A_2 ) > {\cal S} (A_1) + {\cal S}(A_2) $$
which is valid only if the coefficient in front of the log correction is
negative, as in our case. 
It is also interesting to note that under certain assumptions,
when black holes are thought to be composed of mutually non-interacting
but distinguishable 
particles in the {\it grand canonical ensemble} with fluctuating 
particle number, the entropy of the resultant system turns out to be 
proportional to area, followed by log area term \cite{ms},
although in the latter case the log correction 
occurs with a positive sign. 
Also, it might be useful to understand 
the implications of these corrections for the holographic hypothesis.
We hope to report on it in the future. 

\vs{.4cm}
\no
{\bf Acknowledgements}

SD and PM thank G. Kunstatter for useful discussions
and interesting comments. SD thanks S. Carlip and V. Husain for useful
correspondence which included several interesting comments and suggestions,
A. Dasgupta for bringing to his attention refs. \cite{mann} and \cite{by}
and J. Bland for comments.
PM thanks T. R. Govindarajan, R. Kaul, R. B. Mann, M. V. N. 
Murthy and D. Page for helpful comments. He also thanks the 
Physics Departments at the Universities at Winnipeg and Waterloo
for hospitality. 
This work was supported in part by 
the Natural Sciences and Engineering Research Council of Canada.

\end{document}